\documentclass[prl,twocolumn,showpacs]{revtex4}
\usepackage{graphicx}
\usepackage{color}
\usepackage{CJK}

\newcommand{\pa}{\partial}

\newcommand{\td}{\tilde}

\begin{document}
\begin{CJK*}{GB}{gbsn}

\title{Evidence for Special Relativity with de Sitter Space-Time
Symmetry \footnote{Supported by the grants from the NSF of China
under Grant No.~10975128, the grant from 973 Program of China under
Grant No.~2007CB815401.}}

\author{YAN Mu-Lin (скст┴п)\footnote{Email: mlyan@ustc.edu.cn}} \affiliation{Interdisciplinary Center for Theoretical Study,
Department of Modern Physics, University of Science and Technology
of China, Hefei 230026 }


\begin{abstract}
\noindent I show the formulation of de Sitter Special Relativity
(dS-SR) based on Dirac-Lu-Zou-Guo's discussions.  dS-SR quantum
mechanics is formulated, and the dS-SR Dirac equation for hydrogen
is suggested. The equation in the earth-QSO framework reference is
solved by means of the adiabatic approach. It's found  that the
 fine-structure ``constant" $\alpha$ in dS-SR
varies with time. By means of the $t-z$ relation of the $\Lambda$CDM
model,
 $\alpha$'s time-dependency becomes redshift $z$-dependent.
 The dS-SR's predictions of
$\Delta\alpha/\alpha$ agree with data of spectra of 143 quasar
absorption systems, the dS-space-time symmetry is $SO(3,2)$ (i.e.,
anti-dS group) $\;$ and the universal parameter $R$ (de Sitter
ratio) in dS-SR is estimated to be $R\simeq 2.73\times 10^{12}ly$.
The effects of dS-SR become visible at the cosmic space-time scale
(i.e., the distance $\geq 10^9 ly$). At that scale dS-SR  is more
reliable than Einstein SR. The $\alpha$-variation with time is an
evidence of SR with de Sitter symmetry.
\end{abstract}

\pacs{ 03.30.+p; 03.65.Ge; 32.10.Fn; 95.30.Ky; 98.90.+s\\
\vskip0.01in \noindent Keywords: Special Relativity; de Sitter
spece-time symmetry; Quasar; Varying fine-structure constant.}

\maketitle

\section{I, Introduction}

\noindent  Einstein's Special Relativity (E-SR) is the cornerstone
of physics, and any discovery beyond E-SR would be very significant.
E-SR indicates the space-time metric is
$\eta_{\mu\nu}=diag\{+,-,-,-\}$. The most general transformation to
preserve metric $\eta_{\mu\nu}$ is Poincar\'e group. It is well
known that the Poincar\'e group is the limit of the de Sitter group
with sphere radius $R\rightarrow \infty$. Thus people could pursue
whether there exists another type of de Sitter transformation with
$R \rightarrow finite$ which also leads to a Special Relativity
theory (SR). In 1935, P.A.M. Dirac presented an electron wave
equation in de Sitter space, and suggested the study of atomical
physics in the
 equation based on such a kind of special relativity, i.e.,
the Special Relativity with de Sitter symmetry (dS-SR)\cite{Dirac}.
 Differing from General Relativity (GR), SRs rely on two
principles: 1) The inertial motion law for free particle must hold;
2) There must exist a specific space-time symmetry in the
frameworks. Both E-SR and dS-SR satisfy these two principles (see
below). To address the difference between GR and SR, Dirac pointed
out \cite{Dirac} that  the de Sitter space-time is associated {\it
``with no local gravitational fields"} (just like the case in
Minkowski space).

 In this paper I will study dS-SR, and solve dS-SR Dirac equation
of hydrogen atom by means of adiabatic approximation, and show that
the time-variation of fine-structure constant reported by
\cite{webb06,Murphy03b,webb01,webb99} is an evidence for dS-SR,
 and hence an effect beyond E-SR. In other words, the true SR for
 real world is dS-SR with $SO(3,2)$ dS-space-time symmetry (or anti-dS group) and
 dS sphere radius $R\simeq 2.73\times 10^{12} ly$ instead of E-SR.

\begin{figure}[ht]
\includegraphics[scale=0.4]{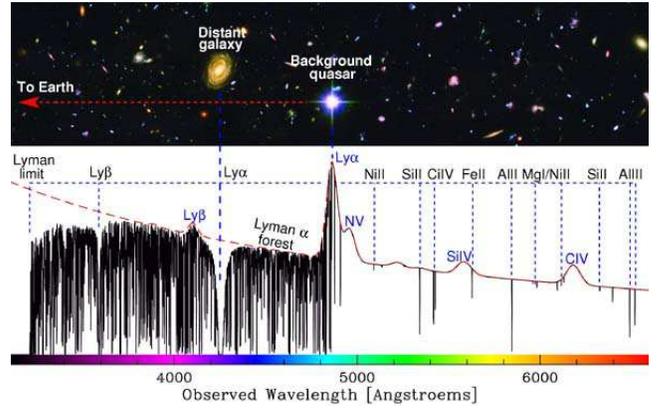}
\caption{\label{fig1} \small A sketch map showing an example of a
spectrum of gas clouds seen in absorption against background
quasi-stellar object (QSO) (download from M.T. Murphy's slide file
(2009)).  }
\end{figure}

Spectroscopic observations of gas clouds seen in absorption against
background quasi-stellar object (QSO) (see Fig 1) have been used to
search for time variation of $\alpha\equiv e^2/(\hbar c)$. Comparing
the observations with the corresponding atomic spectra measured in
laboratory, the results clearly show the first experimental evidence
for the fundamental physics constant variations
\cite{webb06,Murphy03b,webb01,webb99}. Even though there are some
debates on the results \cite{webb06}, this discovery is very
significant, and greatly stimulated the various theoretical
discussions during the last decade (e.g., see
\cite{uzan}\cite{rev2}). BSBM model \cite{Bekenstein,SBM,Barrow} is
one of them. This theory models the variation of $\alpha$ by means
of a scalar field which obeys a Euler-Lagrangian equation derived
from an action. Combining the scalar field theory with General
Relativity (GR) and adjusting the model's parameters, one can get
suitable results describing the $\alpha$-variation
 and evolutions along with $z$ (redshift). However, the price paid for the successes of
 BSBM is that an unknown matter field (i.e., scalar field) has to be
introduced. Some authors called the force propagating by the quanta
of such unknown field as {\it `` fifth force"} \cite{5-force}, which
breaks the electric charger conservation law \cite{22}, and violates
the weak equivalence principle \cite{SBM}. There is not yet any
 experimental evidence to show the existence of such  material scalar
field so far besides explaining time-variation of $\alpha$. In this
case, therefore, searching for alternative scenario without any
unknown particle to explain the $\alpha$-variation with time would
be more conservative, and hence more reliable  for solving the
puzzle.

Moreover, the absorption spectra observations resulting in
declaration of $\alpha$-variation with time reported by
\cite{webb06,Murphy03b,webb01,webb99} rely on the measurements of
the spectrum's fine-structures  of atoms and ions at gas clouds near
QSO. So, if possible, Quantum Mechanics (QM) calculations of atomic
spectra for atoms in the distance  in some suitable model would be a
direct  answer to  the puzzle. For example, the dS-SR atomic physics
scheme suggested by Dirac \cite{Dirac} should be considered
seriously. As is well known, the spectra fine-structures in atomic
physics represent E-SR corrections to levels, which are in principle
derived from E-SR Dirac equation in QM. Especially, E-SR Dirac
equation of hydrogen in QM has exact solution, and the calculations
of such corrections are sound. Those corrections are space-time
independent, and hence $\alpha$ is a constant due to the space-time
translation invariant symmetry of E-SR. Thus, it should be very
interesting to pursue what  the dS-SR corrections to the levels of
atoms in distance are in QM by means of solving the dS-SR Dirac
equation of hydrogen. Because the time translations of dS-SR are
significantly different from those of E-SR, one could expect that
dS-SR QM may yield time-dependent $\alpha$, and lead to solving the
puzzle. In the following, I pursue this topic.

\section{II, Solutions of hydrogen's dS-SR Dirac equation}

In order to precisely  formulate the dS-SR space-time theory and
dynamics, in 1970--1974, LU, ZOU and GUO \footnote{LOOK K H (LU
Qi-Keng), {\it Why the Minkowski metric must be used ?}, (1970),
unpublished.} \cite{Lu74}( for the English version, see Refs.
\cite{Guo} \cite{Yan1}) proved two theorems as follows:

Lemma I: Inertial motion law for free particles holds to be true in
the de Sitter space characterized by Beltrami metric
\begin{eqnarray}\label{star28}
B_{\mu\nu}(x)={\eta_{\mu\nu} \over \sigma (x)}+{{\lambda \;} \over
R^2 \sigma(x)^2} \eta_{\mu\lambda}\eta_{\nu\rho} x^\lambda x^\rho,
\end{eqnarray}
where { $\sigma(x)\equiv 1-{\lambda\over R^2} \eta_{\mu\nu}x^\mu
x^\nu,$ $R^2>0$, and $\lambda =1$ or $-1$ which corresponds to dS
symmetries $SO(4,1)$ or $SO(3,2)$ respectively.} And the constant
$R$ is the radius of the pseudo-sphere in {\it dS}-space. This means
that in dS space characterized by $B_{\mu\nu}$, the velocity of free
particle is constant, i.e.,
\begin{equation}\label{motion}
\dot{\bf{x}}=\overrightarrow{v}=cnstant,~~~~{\rm{for}\;\rm{free\;particle}}
\end{equation}
which is exactly the counterpart  of E-SR's inertial law in
Minkowski space characterized by $\eta_{\mu\nu}$. (see  Refs.
\cite{Guo} \cite{Yan1} for the English version of proof to
Eq.(\ref{motion})).

 Lemma II: The  de Sitter space-time transformation
 preserving $B_{\mu\nu}(x)$ is as follows
\begin{eqnarray}\label{transformation}
x^{\mu}\longrightarrow \tilde{x}^{\mu} &=& \pm \sigma(a)^{1/2}
\sigma(a,x)^{-1}
(x^{\nu}-a^{\nu})D_{\nu}^{\mu}, \\
    \nonumber D_{\nu}^{\mu} &=& L_{\nu}^{\mu}+\lambda R^{-2} \eta_{\nu
\rho}a^{\rho} a^{\lambda} (\sigma
(a) +\sigma^{1/2}(a))^{-1} L_{\lambda}^{\mu} ,\\
\nonumber L : &=& (L_{\nu}^{\mu})\in SO(1,3), \\
\nonumber \sigma(x)= 1&-&{\lambda \over R^2}{\eta_{\mu \nu}x^{\mu}
x^{\nu}},~~  \sigma(a,x)= 1-{\lambda \over R^2}{\eta_{\mu
\nu}a^{\mu} x^{\nu}}.
\end{eqnarray}
where $ x^{\mu}$ is the coordinate in an initial Beltrami frame, and
$ \tilde{x}^{\mu}$ is in another Beltrami frame whose origin is
$a^{\mu}$ in the original one. There are 10 parameters in the
transformations between them. Under the transformation
(\ref{transformation}), we have the equation  preserving
$B_{\mu\nu}$ as follows
\begin{equation} \label{B01}
 B_{\mu\nu}(x)\longrightarrow \widetilde{B}_{\mu\nu}(\widetilde{x})={\pa x^\lambda \over \pa
 \widetilde{x}^\mu}{\pa x^\rho \over \pa
 \widetilde{x}^\nu}B_{\lambda\rho}(x)=B_{\mu\nu}(\widetilde{x}).
\end{equation}
( see  Appendix of Ref. \cite{Yan1} for the English version of proof
to Eq.(\ref{B01})). Eq.(\ref{B01}) will yield conservation laws for
the  energy, momenta, angular momenta and boost chargers of
particles in dS-SR mechanics \cite{Yan1}.

Based on those two lemmas, Yan, Xiao, Huang and Li formulated the
Lagrangian-Hamiltonian formulism for dS-SR dynamics with two
universal constants $c$ and $R$, and the dS-SR Dirac equation has
been proved to be
 \cite{Yan1}\cite{Ut}\cite{NY}\cite{Yan2}:
\begin{equation}\label{Cur-Dirac}
\left(ie_a^\mu \gamma^a D_\mu-{m_0c \over \hbar}\right)\psi=0,
\end{equation}
where $e_a^\mu$ is the tetrad satisfying $e_a^\mu
e_b^\nu\eta^{ab}=B^{\mu\nu}$, and $D_\mu=\pa_\mu-{i\over
4}\omega^{ab}_\mu\sigma_{ab}$ is the covariant derivative with
Lorentz spin connection $\omega^{ab}_\mu$ derived from $B_{\mu\nu}$
of eq.(\ref{star28}). \vskip0.03in \noindent  Furthermore, by gauge
principle, { $D_\mu\rightarrow \mathcal{D}_\mu=D_\mu-ie/(c\hbar
)A_\mu)$ with $A_\mu=B_{\mu\nu}A^\nu$, $A^\nu=(\phi,\;
\mathbf{A}=0)$ } \vskip0.03in \noindent and
$-B^{ij}\pa_i\pa_j\phi={-4\pi e\over
\sqrt{-\det(B_{ij})}}\delta^{(3)}(\mathbf{x})$ where $\phi$ is the
proton's electric Coulomb potential, one has the dS-SR Dirac
equation for the electron in hydrogen atom as follows
\begin{equation}\label{Dirac1}
(ie^\mu_a \gamma^a\mathcal{D}_\mu^L -{\mu c\over \hbar })\psi=0,
\end{equation}
where $$\mu=m_e/(1+{m_e\over m_p})$$ is the reduced mass of
electron. In this formulism, the  measurable conserved 4-momentum
operator is \cite{Yan1}
\begin{eqnarray}\label{momentum operator}
p^{\mu} &=& i\hbar \left[\left(\eta^{\mu\nu}-\frac{\lambda
x^{\mu}x^{\nu}}{R^2}\right)\partial_{\nu}+\frac{5\lambda
x^{\mu}}{2R^2}\right].
\end{eqnarray}

The observation results reported by
\cite{webb06,Murphy03b,webb01,webb99} are the absorption spectra of
gas clouds against background QSO. We briefly call the gas-QSO
system as QSO for simplicity. We are interested in the atoms,
typically the hydrogen atom, at QSO that  locates on the light-cone
in de Sitter space with Beltrami metrics because only this kind of
QSO can be observed by earth-observers.  As illustrated in Fig.2(a),
the earth locates at the origin of the frame, the proton (nucleus of
hydrogen atom) locates at $Q=\{Q^0\equiv
c\;t,\;Q^1=c\;t,\;Q^2=0,\;Q^3=0\}$, which is on QSO-light-cone
$B_{\mu\nu}(Q)Q^\mu Q^\nu=\eta_{\mu\nu}Q^\mu Q^\nu=0$. The metric of
the space-time near $Q$ is
$B_{\mu\nu}(Q)=\eta_{\mu\nu}+{\lambda \over R^2}\eta_{\mu \lambda
}Q^\lambda\eta_{\nu \rho}Q^\rho,$ and hence
$B_{ij}(Q)=\eta_{ij}+{\lambda c^2 t^2\over R^2}
\delta_{i1}\delta_{j1}.$
 Electron-coordinates are $L=\{L^0\equiv
c t_L,\;L^1,\;L^2,\;L^3\}$, and the relative space coordinates
between proton and electron are $x^i=L^i-Q^i$. The magnitude of
$r\equiv \sqrt{-\eta_{ij}x^ix^j}\sim a$ (where $a\simeq 0.5\times
10^{-10}m$ is Bohr radius), and $|x^i|\sim a$. Another scale is the
Compton wave length of electron $a_c=\hbar/(m_e c)\simeq 0.3\times
10^{-12}m$. Noting $R$ is cosmologically large and $R>>ct$, so the
calculations for our purpose will be accurate up to
$\mathcal{O}(c^2t^2/R^2)$. The terms proportional to
$\mathcal{O}(c^4t^4/R^4)$, $\mathcal{O}(cta_c/R^2)$,
$\mathcal{O}(cta/R^2)$, etc. will be ignored. Note also that
Eq.(\ref{momentum operator}) indicates that the energy eigenstate
equation is
$$
 E\psi=i\hbar\left[\pa_t-{\lambda c^2t^2\over
R^2}\pa_t+{5\lambda ct\over 2R^2}\right]\psi\simeq
i\hbar\left(1-{\lambda c^2t^2\over R^2}\right)\pa_t\psi.
$$
Then Eq.(\ref{Dirac1}) becomes {
\begin{eqnarray}
\nonumber E\psi&=&\left[ -i\hbar c\hskip-0.05in\left(1-{\lambda
c^2t^2\over 2R^2}\right) \vec{\alpha}\cdot\nabla_B
\hskip-0.06in+\hskip-0.06in\left(1-{\lambda c^2t^2\over 2R^2}\right)
\mu
c^2\beta \right.\\
\label{Dirac7}&& -\left.{{e}^2\over r_B} \right]\psi,
\end{eqnarray} }
\vskip0.02in \noindent where
$r_B=\sqrt{(\td{x}^1)^2+(x^2)^2+(x^3)^2}$ with $\td{x}^1=(1-\lambda
c^2t^2/(2R^2))x^1$ and $\nabla_B = \mathbf{i}{\pa \over
\pa\td{x}^1}+ \mathbf{j}{\pa \over \pa x^2}+\mathbf{k}{\pa\over \pa
x^3}$. Eq.(\ref{Dirac7}) is time-dependent quantum Hamiltonian
equation. It is somewhat difficult to deal with the time-dependent
 problems in quantum mechanics.  Fortunately, comparing (\ref{Dirac7}) with usual E-SR Dirac equation
for hydrogen, all correction terms duo to dS-SR are  proportional to
$(c^2t^2/R^2)$. Since $R>>ct$, those factors make the time-evolution
of the system  so slow that the adiabatic approximation \cite{Born}
will legitimately work (see Chapter XVII of Vol II of
\cite{Messian}, and Appendix B in \cite{Yan2}). Thus, rewriting
(\ref{Dirac7}) as $ E\psi=\left[ -i\hbar_t c
\vec{\alpha}\cdot\nabla_B + \mu_t c^2\beta -{{e_t}^2\over r_B}
\right]\psi$ with \vskip0.03in $\hbar_t= \left(1-{\lambda
c^2t^2\over 2R^2}\right)\hbar,\; \mu_t= \left(1-{\lambda c^2t^2\over
2R^2}\right)\mu,\; { e_t =e, }$ we obtain the predictions {
\begin{equation}
\label{alpha} \alpha_t \equiv {e_t^2\over \hbar_t c}=(1+{\lambda
c^2t^2\over 2R^2}){\alpha},~~{\rm or}~~ {\Delta\alpha\over
\alpha}={\lambda c^2t^2\over 2R^2},
\end{equation}}
\begin{eqnarray}
\nonumber \omega_t&=&E/\hbar_t={\mu_t\over \hbar_t}
c^2\left[1+{\alpha_t^2
\over (\sqrt{K^2-\alpha_t^2}+n_r)^2} \right]^{-1/2} \\
\label{solution2} &=&{\mu\over \hbar} c^2\left[1+{\alpha_t^2 \over
(\sqrt{K^2-\alpha_t^2}+n_r)^2} \right]^{-1/2}.
\end{eqnarray}
\begin{figure}[ht]
\includegraphics[scale=0.30]{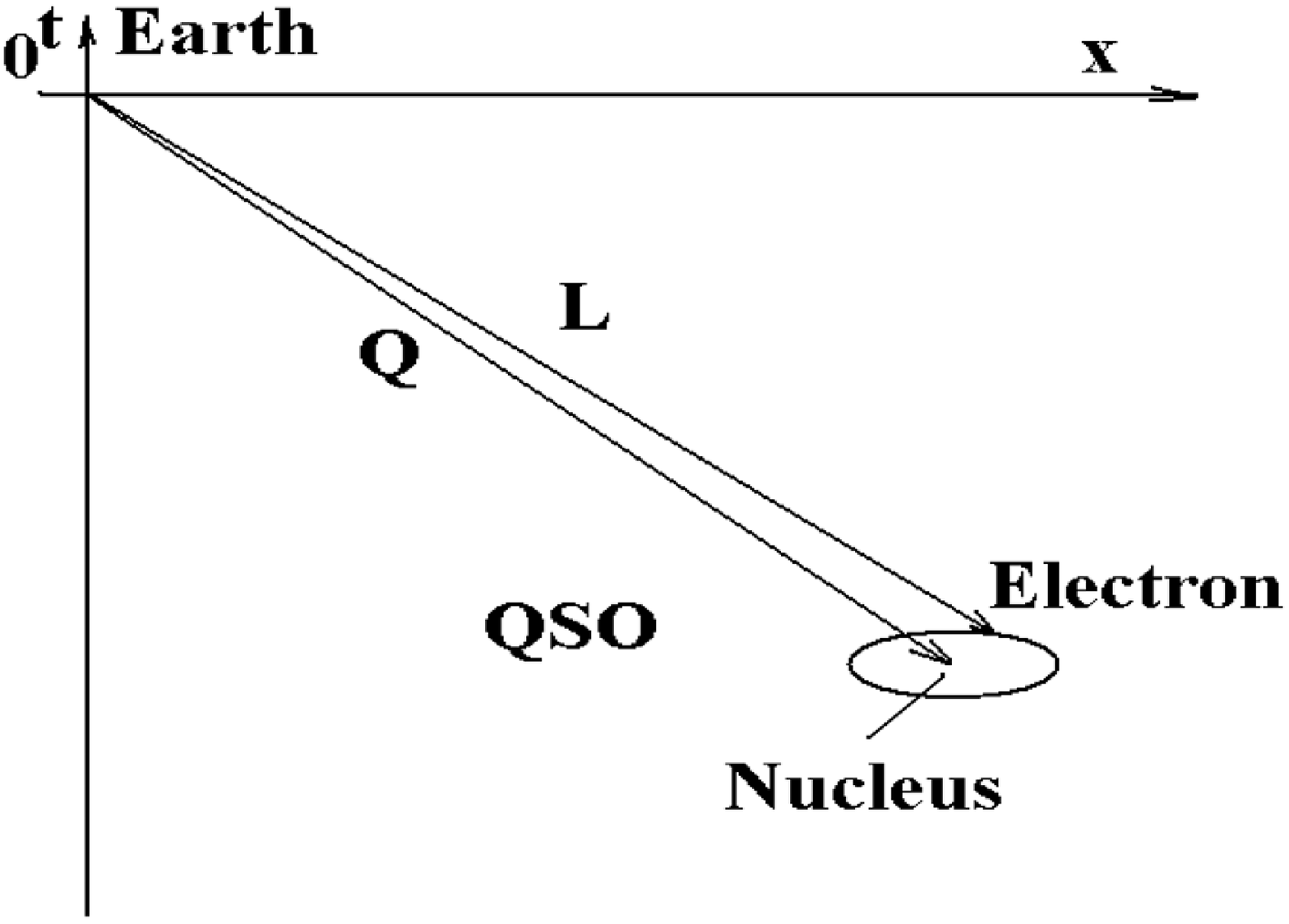}
\caption{\label{fig2} \small Sketch of the earth-QSO reference
frame. The earth locates at the origin. The position vector for the
nucleus of the atom on QSO is $Q$, and for the electron is $L$. The
distance between the nucleus and  electron is $r$. }
\end{figure}

\section{III, Comparison between theory predictions and observation
data}

Murphy and collaborators \cite{Murphy03b} studied the spectra of 143
quasar absorption systems over the redshift range $0.2<
z_{abs}<4.2$. Their most robust estimate is a weighted mean
\begin{equation} \label{ex1}
{\Delta \alpha \over \alpha}=(-0.57\pm 0.11)\times 10^{-5}.
\end{equation}
{ Compared  with the prediction (\ref{alpha}), we conclude that
\begin{equation} \label{lambda}
\lambda=-1.
\end{equation}
 This means that the space-time symmetry for dS-SR is de
Sitter-$SO(3,2)$ instead of anti-de Sitter-$SO(4,1)$. Substituting
Eq.(\ref{lambda}) into (\ref{alpha}), we predict as follows
\begin{equation} \label{alpha1}
{\Delta\alpha\over \alpha}=-{ c^2t^2\over 2R^2}.
\end{equation}}

The 134 data points are assigned three epochs in Ref. \cite{Dent}
(see Table I), and the redshift $z$-dependence of $\Delta
\alpha/\alpha$ is shown roughly in \cite{Dent}. In the following, I
further test the prediction of (\ref{alpha}) in terms of these
$z$-dependent data of $\Delta \alpha/\alpha$. In order to transfer
the $t$-dependence of $\Delta \alpha/\alpha$ in (\ref{alpha}) to a
$z$-dependence prediction,  a relation of $t-z$ is needed. For this
aim, an appropriate cosmological model is necessary. In this paper,
we treat $t$ as  comoving time $t$ in the $\Lambda$CDM model
\cite{Lambda,Lambda1}. In the model, the $t-z$ relation is as
follows
\begin{equation}\label{La1}
t=\int_0^z{dz' \over H(z')(1+z')},
\end{equation}
where
\begin{eqnarray*}
\label{La2}H(z')&=&H_0\sqrt{\Omega_{m0}(1+z')^3+1-\Omega_{m0}},\\
\label{La3}H_0&=&100\;h\simeq 100\times0.705 km\cdot s^{-1}/Mpc,\\
\label{La4}\Omega_{m0}&\simeq &0.274.
\end{eqnarray*}
The $t-z$ relation is shown in Fig.3(a). Substituting this relation
into (\ref{alpha}), we obtain a desirable $z$-dependence prediction
of ${\Delta  \alpha \over \alpha} (z)$, where $R$ is a free
parameter. By using the observation data ${\Delta  \alpha \over
\alpha} (z=1.47)=-0.58\times 10^{-5}$, we get $R\simeq 2.73\times
10^{12} ly$ (which is consistent with the estimation in
\cite{Yan3}). Then the theory predictions are ${\Delta  \alpha \over
\alpha} (z=0.65)=-0.24\times 10^{-5}$ and ${\Delta  \alpha \over
\alpha} (z=2.84)=-0.87\times 10^{-5}$, which are in agreement with
the corresponding data in \cite{Murphy03b} and \cite{Dent}. The
results are listed in Table I, and the curve of ${\Delta  \alpha
\over \alpha} (z)$ is shown in Fig.3(b). The comparison indicates
that the dS-SR theory predictions of (\ref{alpha}) agree with the
observation data within the error band.
\begin{figure}[ht]
\includegraphics[scale=0.5]{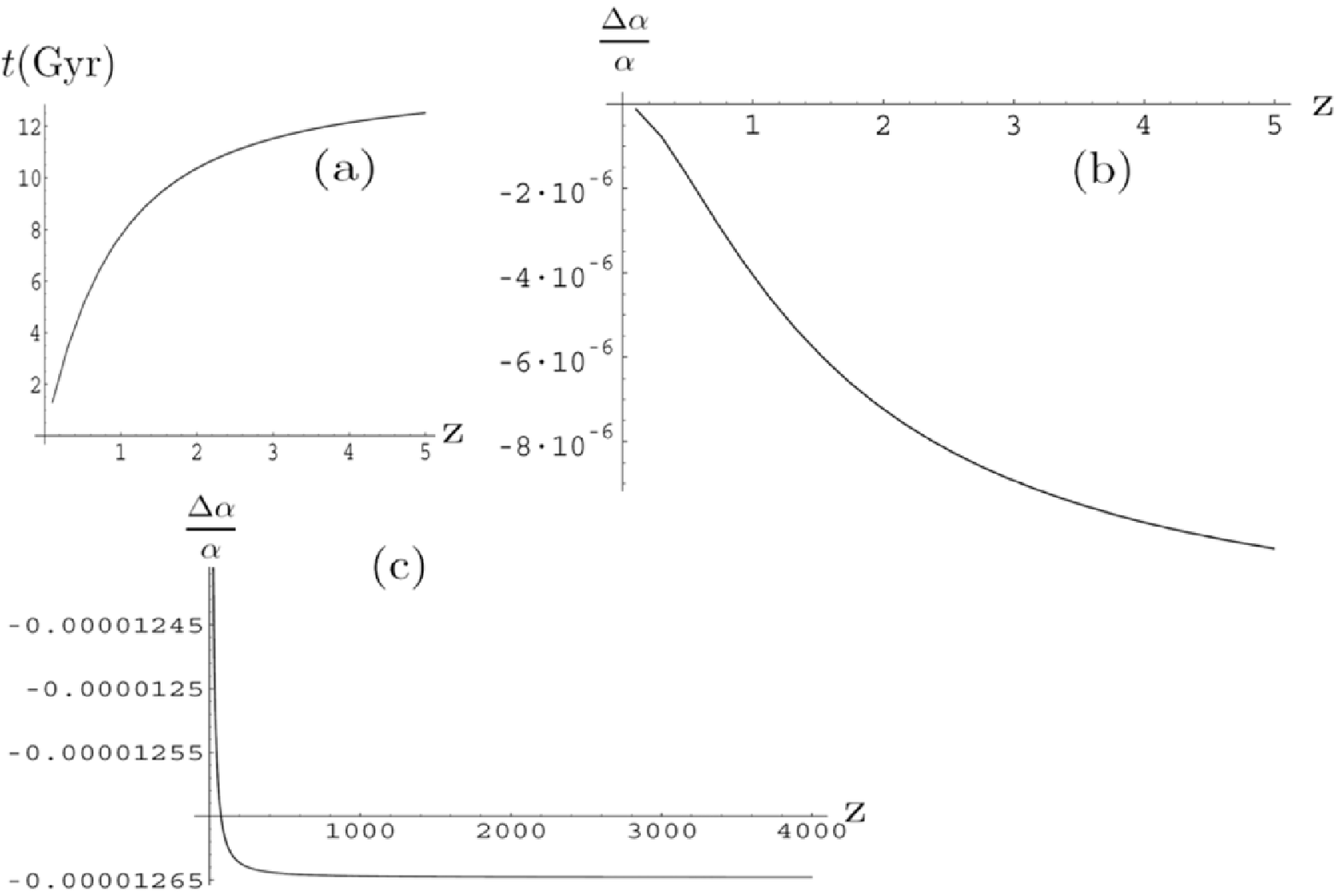}
\caption{\label{fig} \small (a) The $t-z$ relation in $\Lambda$CDM
model; (b) The $\Delta \alpha/\alpha$ as function of the red shift
$z$; (c) The evolution of $\alpha$-variations
${\Delta\alpha\over\alpha}(z)$ along with $z$. By Eq. (\ref{alpha1})
 with $R\simeq 2.73\times 10^{12} ly$ and the
$\Lambda$CDM model's $t-z$ relation,  ${\Delta\alpha\over\alpha}(z)$
curve is plotted in the region of $(0\leq z \leq 4000)$.}
\end{figure}
\begin{table}
\caption{ Time variations of $\Delta \alpha/\alpha$: The first two
columns are quoted from \cite{Dent}. Eq. (\ref{alpha1})
 with
 $R\simeq 2.73\times 10^{12} ly$, and the $\Lambda$CDM model's $t-z$ relation (\ref{La1}) are used. }
\tabcolsep 0.1in
\begin{tabular}{l l l} \hline \hline
redshift  $\langle z\rangle$ and ($t$)& $(\Delta
\alpha/\alpha)_{expt}$ &  results of (\ref{alpha1}) \\
\hline 0.65 ($6.04$Gyr) & $(-0.29\pm0.31)\times 10^{-5}$
&$-0.24\times 10^{-5}$
\\
1.47 ($9.29$Gyr)& $(-0.58\pm0.13)\times 10^{-5}$ &$-0.58\times
10^{-5}$
\\
2.84 ($11.39$Gyr) & $(-0.87\pm0.37)\times 10^{-5}$ &$-0.87\times
10^{-5}$
\\\hline \hline
\end{tabular}
\end{table}

Next, we turn to discuss the evolution of $\alpha$-variations
${\Delta\alpha\over\alpha}(z)$ along with $z$, and plot
${\Delta\alpha\over\alpha}(z)$ curve in the region of $(0\leq z \leq
4000)$ in Fig.3(c). We can see that as $z<10$,
${\Delta\alpha\over\alpha}(z)$ changes relatively sharply, and then
the changes become slow. When $z\geq 10^3$,
 ${\Delta \alpha\over \alpha}(z) $ is almost independent of $z$,
 i.e.,  $\alpha$-variation ceases in that
 very high $z$ region. Fig 3(c) shows that
 the lower bound of ${\Delta \alpha\over \alpha}(z) $ is about $\sim -1.3\times
 10^{-5}$. This result coincides with other  considerations (e.g., BSBM
 model)
 \cite{SBM}, which suggests a negligible change in
 $\alpha$ in the radiation epoch of the Universe, that epoch roughly corresponds to $z\geq 3\times 10^{3}$.

\section{IV, Conclusion}

In summary, in this paper, I have shown the formulation of de Sitter
Special Relativity (dS-SR) based on Dirac-Lu-Zou-Guo's discussions,
formulated the dS-SR quantum mechanics, and then determined  the
dS-SR Dirac equation for hydrogen. In order to discuss the spectra
of atoms on (or near) QSO,  I solved it in the earth-QSO framework
reference by means of the adiabatic approach. Aspects of de Sitter
space-time geometry described by Beltrami metric are taken into
account. The dS-SR Dirac equation of hydrogen turns out to be
 a time dependent quantum Hamiltonian system.
 Since the radius
of de Sitter sphere $R$ is cosmologically large, it  makes the
time-evolution of the system so slow that the adiabatic
approximation legitimately works with high accuracy. Consequently,
it is revealed that all those facts yield important conclusions that
the electromagnetic fine-structure ``constant" $\alpha$ varies with
time. By means of the $t-z$ relation of the $\Lambda$CDM model, the
$\alpha$'s time-dependent becomes redshift $z$-dependent. The
dS-SR's predictions of $\Delta\alpha/\alpha$
 are in agreement with the data,  the
dS-space-time symmetry is $SO(3,2)$ (i.e., anti-dS group) and the
universal parameter $R$ (the de Sitter ratio) in the theory is
estimated to be $R\simeq 2.73\times 10^{12}ly$. This fact indicates
that the effects of dS-SR become visible at the cosmic space-time
scale (i.e., the distance $\geq 10^9 ly$).  At that scale de Sitter
Special Relativity is more reliable than Einsteinian Special
Relativity, and the latter is the former's approximation for the
distance which is much less than $ R$, or much less than $\sim
10^9ly$. I conclude that the $\alpha$-variation with time is
evidence of SR with de Sitter symmetry.

\vskip-0.2in

{

}
\end{CJK*}

\begin{thebibliography}{99}

\bibitem{Dirac} Dirac P A M, Annals of Mathmatics, 1935, {\bf 36}(3):
657-669
\bibitem{webb06}
Murphy M T, Webb J K, Flambaum V V,  Phys. Rev. Lett., 2007, {\bf
99}:  239001-239001 arXiv:0708.3677 [astro-ph] ; astro-ph/0612407;
astro-ph/0911.4512

\bibitem{Murphy03b}
Murphy M T,  Flambaum V V, Webb J K, Dzuba V V, Prochaska J X, Wolfe
 A M, Lec. Notes in Phys., 2004, {\bf 648}:  131



\bibitem{webb01}
 Webb J K,  Murphy M T,  Flambaum V V, Dzuba V A, Barrow J D,
Churchill C W, Pochaska J X, Wolfe A M, { Phys. Rev. Lett}., 2001,
{\bf87}: 091301-091304.

\bibitem{webb99}
Webb J K, Flambaum V V, Churchill C W, Drinkwater M J, Barrow J D,
Phys. Rev. Lett., 1999, {\bf82}: 884-888.

\bibitem{uzan} Uzan J P, Rev. of Mod. Phys., 2003 {\bf 75}: 403-455
\bibitem{rev2} Flambaum V V, Int. J. Mod. Phys., 2007, {\bf A 22}: 4937-4950

\bibitem{Bekenstein} Bekenstein J D, Phys. Rev, 1982, {\bf D25}:
1527-1539

\bibitem{SBM} Sandvik H B, Barrow J D, and Magueijo J, Phys. Rev.
Lett., 2002 {\bf 88}: 031302-1--031302-4
\bibitem{Barrow} Barrow J D, Phys Rev., 2005, {\bf D71}: 083520-1--083520-7;  Barrow J D, Sandvik H B, and Magueijo J, Phys. Rev.
, 2002, {\bf D65}: 063504-1--063504-9; {\it ibid}, 2002, {\bf D66}:
043515-1--043515-6

\bibitem{5-force} Dvali G and  Zaldarriaga M, Phys Rev Lett.,
2002, {\bf 88}: 091303-1--091303-4
\bibitem{22} Landau S, Sisterna P, and Vucetich H, Phys. Rev., 2001, {\bf D
63}: 081303(R)-1--081303-4


\bibitem{Lu74} LOOK K H (LU Qi-Keng), TSOU C L (ZOU Zhen-Long) and KUO H Y (
GUO Han-Ying),  Acta Physica Sinica, 1974 {\bf 23}: 225-237, (in
Chinese)

\bibitem{Guo} GUO Han-Ying, HUANG Chao-Guang, XU Zhan and ZHOU Bin, Phys. Lett., 2004,  {\bf A 331}:
1-7, hep-th/0403171

\bibitem{Yan1} YAN Mu-Lin, XIAO Neng-Chao, HUANG Wei, LI Si, Commun. Theor.
Phys., 2007, {\bf 48}: 27-36, hep-th/0512319.

\bibitem{Ut} Utiyama R, Phys. Rev., 1956, {\bf 101}: 1597-1607
\bibitem{NY} NIEH H T and  YAN Mu-Lin, Ann. Phys., 1982, {\bf 138}: 237-259, and the
references within.
\bibitem{Yan2} YAN Mu-Lin, Int. Mod. Phys A (to appear), {\it ``Hydrogen Atom in de Sitter Special Relativity  and Time Variation of
Fine-Structure Constant"}, arXiv: 1004.3023 [physics.gen-ph].

\bibitem{Born} Born M and Fock V, Z. Phys., 1928, {\bf 51}, 165-175
\bibitem{Messian} Messiah A, Quantum Mechanics I, II. North-Holland
Publishing Company, 1970.


\bibitem{Dent}  Dent T, Stern T, Wetterich C, Phys. Rev. {\bf D78}
(2008) 103518-1--103508-17.
\bibitem{Lambda} Weinberg S, Rev. Mod. Phys., 1989, {\bf 61}: 1-23; Padmanabhan T, Phys. Rep., 2003, {\bf 380}: 235-320

\bibitem{Lambda1}  Komatsu E, {\it et al},
Astrophys.J.Suppl., 2009, {\bf 180}: 330-376

\bibitem{Yan3} CHEN Shao-Xia,  XIAO Neng-Chao,  YAN Mu-Lin, Chinese Physics, 2008, {\bf
C32}: 612-616, astro-ph/0703110.





\end{thebibliography}
\end{document}